# Theory-Guided Discovery of Pressure-Induced Transitions in Fast-Ion Conductor BaSnF$_4$

## Authors

Robin Turnbull[1,*], Zhang YingLong[2], Claudio Cazorla[3], Akun Liang[4,1], Rahman Saqib[2], Miriam Peña-Alvarez[5], Catalin Popescu[6], Laura Pampillo[7,8], Daniel Errandonea[1]

## Affiliations

*robin.turnbull@uv.es

1 Departamento de Física Aplicada-ICMUV, MALTA Consolider Team, Universidad de Valencia, 46100 Burjassot, Valencia, Spain

2 School of Physics and Optoelectronic Engineering, Hangzhou Institute for Advanced Study, University of Chinese Academy of Sciences, UCAS, China

3 Departament de Física, Universitat Politècnica de Catalunya, Campus Diagonal-Besòs, Barcelona, Spain

4 Eastern Institute of Technology, Ningbo 315200, China

5 Centre for Science at Extreme Conditions and School of Physics and Astronomy, University of Edinburgh, Edingurgh, EH9 3FD, Scotland, United Kingdom

6 CELLS-ALBA Synchrotron Light Facility, Cerdanyola, Barcelona, 08290, Spain

7 Universidad de Buenos Aires, Facultad de Ingeniería, Laboratorio de Sólidos Amorfos, Paseo Colón 850 (1063), Buenos Aires, Argentina

8 CONICET – Universidad de Buenos Aires, Instituto de Tecnologías y Ciencias de la Ingeniería "Hilario Fernández Long" (INTECIN), Buenos Aires, Argentina

**Abstract**

Fast-ion conductors such as $BaSnF_4$ are of significant interest for next-generation solid-state battery technologies due to their high ionic conductivity and chemical stability. However, the behaviour of these materials under extreme conditions remains poorly understood, despite the relevance of pressure-induced modifications for tuning functional properties. In this study, we combine density functional theory (DFT) calculations with high-pressure experiments to investigate the structural evolution of $BaSnF_4$ up to 40 GPa. DFT predicts two pressure-induced phase transitions: from the ambient-pressure tetragonal *P*4/*nmm* phase to a monoclinic *P*2$_1$/*m*-I structure at 10 GPa, and subsequently to a denser monoclinic *P*2$_1$/*m*-II phase at 32 GPa. The first transition is experimentally confirmed via angle-dispersive X-ray diffraction, Raman spectroscopy, and electrical resistivity measurements, all performed at ambient temperature. The second transition is supported by distinct changes in high-pressure Raman modes and resistivity behaviour, consistent with a further structural reorganization. These findings not only clarify the high-pressure phase diagram of $BaSnF_4$, but also shed light on the potential for pressure-tuned ionic transport in fluorostannate-based solid electrolytes.

# I. Introduction

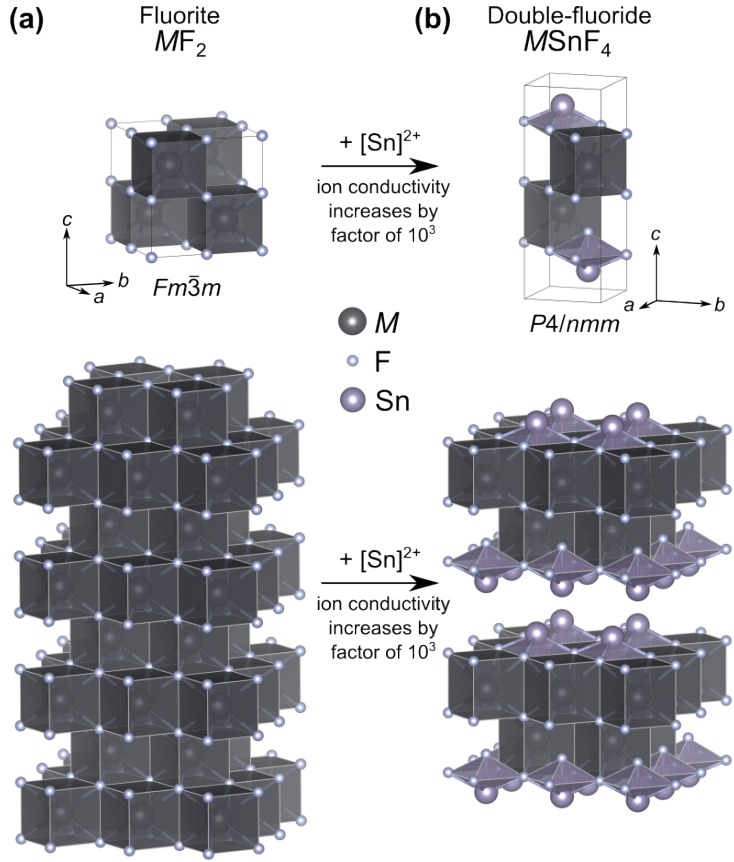

**Figure 1 (a)** The fluorite (*M*F$_2$) structure. **(b)** The *double-fluoride* (*M*SnF$_4$) structure.

Fast-ion conductors are a class of solid-state materials that exhibit a sharp increase in ionic conductivity above a critical temperature [1]. Ion mobility is typically favoured for small, monovalent ions, such as fluoride (F$^-$), which are amongst the most mobile in the solid state. Fluoride-based materials with the fluorite structure type (MF$_2$, where M = Ba, Ca, Sr, Cd, Hg, Eu) are particularly well-studied, with conductivities exceeding 0.1 S cm$^{-1}$ at high temperatures [2]. These structures adopt the $Fm\overline{3}m$ space group and consist of a face-centred cubic (fcc) metal lattice coordinated by eight fluoride ions, while the F$^-$ ions occupy tetrahedral sites coordinated by four metal cations.

Substituting a portion of the M$^{2+}$ ions with Sn$^{2+}$ leads to the formation of double-fluoride compounds with the formula MSnF$_4$. These materials adopt derivatives of the fluorite structure with layered ordering of M and Sn cations, and exhibit significantly enhanced fluoride-ion conductivity, i.e. up to three orders of magnitude higher than the parent MF$_2$ compounds [3]. Their relatively high conductivity, low cost, and absence of flammable liquid electrolytes make them promising candidates for solid-state fluoride-ion batteries, which are of increasing interest as alternatives to lithium-ion technology [4].

Despite their potential, the structural response of double-fluorides to high pressure remains poorly understood. Pressure is known to influence ion transport by modifying the lattice dynamics and the local environment of mobile ions, with theoretical studies suggesting that it may reduce the superionic transition temperature in fluoride-based conductors [5,6]. Moreover, pressure-induced structural transitions in fast-ion conductors are of interest for solid-state cooling applications, where coupling between ionic conductivity and entropy changes may be exploited [7,8].

Here, we investigate the high-pressure behaviour of $BaSnF_4$, a representative double-fluoride fast-ion conductor, using a combination of density functional theory (DFT) calculations and high-pressure experiments. Although $BaSnF_4$ is known to adopt a tetragonal structure (space group *P*4/*nmm*) at ambient conditions and exhibit good ionic conductivity, no experimental data exist regarding its behaviour under compression. Our study aims to clarify the structural phase diagram of $BaSnF_4$ at high pressure and to assess the potential for pressure to influence its transport properties.

**Methods**

**Sample preparation**

BaSnF$_4$ was prepared via a chemical co-precipitation synthesis method inspired by Ref. [9]. Stoichiometric amounts of barium chloride (BaCl$_2$) and stannous chloride (SnCl$_2$) were mixed together and dissolved in distilled water. This solution was then added dropwise into a solution of ammonium fluoride (NH$_4$F). The subsequent chemical reaction precipitates the final product, which is then filtered and dried:

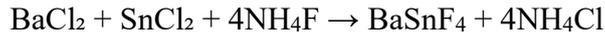

$$BaCl_2 + SnCl_2 + 4NH_4F \rightarrow BaSnF_4 + 4NH_4Cl$$

**High-pressure angle-dispersive X-ray diffraction** was measured using a diamond-anvil cell (DAC) with culets of 500 μm in diameter. A 200 μm thick stainless steel sheet was used as a gasket material. It was indented to a thickness of 40 μm by the diamonds, and then a 150 μm hole was drilled in the centre. An example sample loading is shown in **Supplementary Figure 1**. The pressure medium was a mixture of methanol and ethanol in a 4:1 ratio [10]. Pressure was remotely controlled via a gas-membrane. The experiment was performed at ALBA synchrotron on the BL04-MSPD beamline [11] using an X-ray wavelength of 0.4642 Å (26.7 keV) and beam size of approximately 20 × 20 μm FWHM. The XRD patterns were acquired on a Rayonix SX165 CCD detector at a distance of 240.08 mm, using LaB$_6$ as a standard calibrant. Acquisition time was typically 30 seconds. The sample pressure was measured before and after XRD acquisition using the online ruby system of BL04. The 2D diffraction images were integrated DIOPTAS [12] and Rietveld refinements were performed using Profex [13].

**High-pressure Raman spectra** were acquired using the 514 nm emission line of an Ar$^+$ ion laser at around 25 mW. A Princeton Instruments Acton SpectraPro 2500i spectrometer was utilized, featuring interchangeable diffraction gratings with 300 and 1800 lines/mm. The CCD detector used for the Raman measurements was a Princeton Instruments model 7500-0003. Pressure was determined using the ruby fluorescence (< 25 GPa) and diamond edge scales (> 25 GPa) [14,15]. The Rayleigh scattering contribution to the spectrum is removed by passing the signal through two holographic notch filters. Helium was used as the pressure-transmitting medium and loaded into a piston-cylinder-type DAC equipped with 250 μm culet diamonds and a rhenium gasket, which was indented to a thickness of 28 μm with a 100 μm diameter hole as the sample chamber.

**High-pressure resistivity measurements** were performed using the standard four-probe technique in a DAC with diamond culets of 300 μm in diameter. The sample resistivity was calculated according to the van der Pauw formula. A mixture of epoxy and cubic boron nitride was utilized as the coating on the steel gaskets to ensure electrical insulation between different electrodes. Four platinum electrodes and copper wires were set up to contact the sample in the chamber. No pressure medium was used for these measurements. Pressure was determined using the ruby scale.

**Density Functional Calculations**

First-principles calculations based on density functional theory (DFT) [16] were carried out with the PBEsol exchange-correlation energy functional [17] as it is implemented in the VASP software [18]. The projector-augmented wave method (PAW) [19] was employed to represent the ionic cores by considering the following electronic states as valence: Ba 5s 5p 6s; Sn 5s 5p; F 2s 2p. An energy cutoff of 750 eV and dense Monkhorst-Pack k-point densities (e.g., a 14×14×7 grid for the 12-atom bulk tetragonal unit cell) were used for integrations within the Brillouin zone, leading to total energies converged to within 1 meV per atom. Atomic relaxations were concluded when the forces in the atoms were all below 0.005 eV/Å. Phonon calculations were performed with the small-displacement method [20], using large 3×3×2 supercells containing a total of 216 atoms.

## II. Results and Discussion

**Density functional theory calculations**

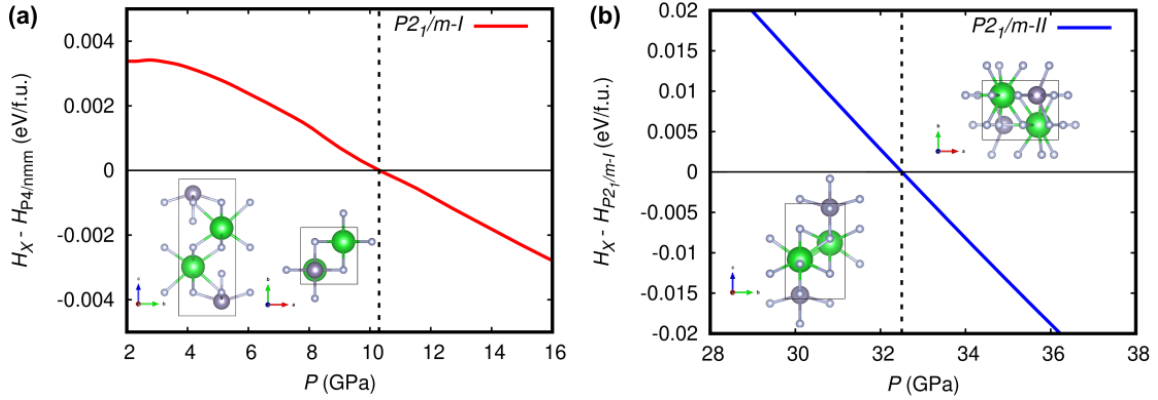

**Figure 2 (a)** Calculated enthalpy of the high-pressure monoclinic ($P2_1/m$-I) BaSnF$_4$ structure relative to the ambient pressure fluorite-type tetragonal structure ($P4/nmm$). **(b)** Calculated enthalpy of the higher-pressure monoclinic ($P2_1/m$-II) BaSnF$_4$ structure relative to the lower-pressure monoclinic ($P2_1/m$-I) BaSnF$_4$ structure of figure 1a. The overall sequence of phase transitions predicted by DFT is: $P4/nmm$ (ambient pressure) → $P2_1/m$-I (>10 GPa) → $P2_1/m$-II (>32 GPa)

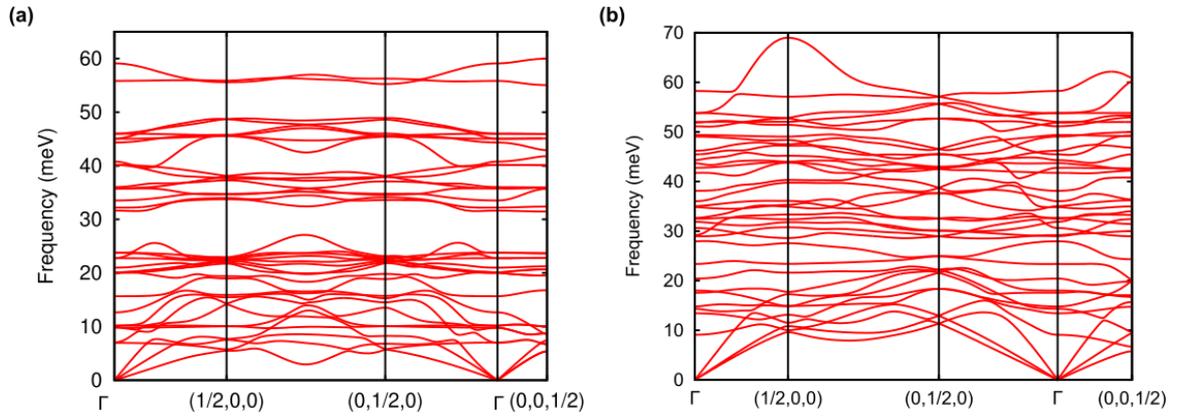

**Figure 3** Calculated phonon frequencies for **(a)** monoclinic ($P2_1/m$-I) BaSnF$_4$ at $P$ = 11 GPa, and **(b)** monoclinic ($P2_1/m$-II) BaSnF$_4$ at $P$ = 34 GPa.

Motivated by the lack of experimental studies on the high-pressure behaviour of double fluorides, we performed density functional theory (DFT) calculations on the double-fluoride $BaSnF_4$ (barium tetrafluorostannate) in order to guide potential high-pressure experiments. Our DFT calculations predicted an ambient-temperature phase transition from the known ambient-pressure fluorite-type tetragonal structure (*P*4/*nmm*) to a monoclinic structure (*P*2$_1$/*m*-I) at just above $P$ = 10 GPa. According to the DFT calculations, the tetragonal $BaSnF_4$ structure is vibrationally unstable at pressures higher than 22.1 GPa. In particular, we observe two Γ phonon modes exhibiting imaginary frequencies at such pressures, suggesting a dynamic instability which often precedes structural phase transition. Such vibrational instabilities are critical in fast-ion conductors, where structural rearrangements can alter ion migration pathways and impact ionic conductivity, as seen in similar fluoride systems [1, 6]. The new low-enthalpy monoclinic phase in this work was determined by introducing the atomic displacements corresponding to one of the two Γ imaginary phonon modes in the unit cell of the tetragonal phase, and letting the system to relax so as to minimise its energy. The tetragonal (*P*4/*nmm*) to monoclinic (*P*2$_1$/*m*-I) phase transition is illustrated in the enthalpy plot in **Figure 2a**.

The calculated phonon spectrum of the high-pressure monoclinic (*P*2$_1$/*m*-I) $BaSnF_4$ structure, which confirms its predicted vibrational stability, is shown in **Figure 3a** calculated at $P$ = 3 GPa. The predicted tetragonal to monoclinic transition was later confirmed experimentally in this work by high-pressure angle-dispersive synchrotron X-ray diffraction HP-XRD (see next section). The calculated unit cell parameters are compared to the experimentally determined unit cell parameters from HP-XRD in the next section, showing good agreement. For the tetragonal phase, the maximum relative error occurs along the c-axis at ambient pressure, reaching approximately 4%. As pressure increases, this error rapidly decreases to below 1%. In the monoclinic phase, the largest relative error is around 3%, observed at the highest applied pressure.

The DFT calculations also predicted a second pressure induced phase transition from the aforementioned monoclinic *P*2$_1$/*m*-I structure to a second monoclinic *P*2$_1$/*m*-II structure, herein called monoclinic *P*2$_1$/*m*-II, above $P$ = 32 GPa (see **Figure 2b**). This second low-enthalpy monoclinic phase was determined by considering the second of the two imaginary Γ phonon modes found for the tetragonal (P4/nmm) phase. Whilst the lower pressure monoclinic *P*2$_1$/*m*-I structure was confirmed by HP-XRD, XRD data have not been acquired at such high pressures, therefore the predicted *P*2$_1$/*m*-II phase cannot be confirmed unambiguously in the present work. However, experimental high-pressure Raman spectra acquired up to $P$ = 40 GPa (see Raman section) and resistivity measurements up to 56 GPa do indicate a possible phase transition around $P$ = 30 GPa (see Resistivity section) in accordance with the DFT calculations.

To quantify the effect of the exchange-correlation functional on the DFT results, the two previously predicted phase transitions were recalculated using the PBE [21], RSCAN [22], and PBEsol-Sn(4d) functionals. The latter corresponds to the standard PBEsol functional, but includes the 4d orbitals of Sn in the valence configuration. As shown in **Table 1**, the calculated transition pressures are consistent across all cases, with particularly close agreement observed between the PBEsol and PBEsol-Sn(4d) results. The PBEsol functional was therefore used to perform most calculations in the present work because it is computationally affordable (especially compared to the PBEsol-Sn(4d) case) and has been shown to be especially good at predicting structural parameters of materials [23-25].

**Table 1** Summary of the phase transition pressures calculated with different exchange functionals.

| $E_{xc}$ functional | $P4/nmm \rightarrow P2_1/m$-I (GPa) | $P2_1/m$-I $\rightarrow P2_1/m$-II (GPa) |
|---|---|---|
| PBEsol | 10.3 | 32.5 |
| PBEsol-Sn(4d) | 10.3 | 33.5 |
| PBE | 13.9 | 36.4 |
| RSCAN | 11.7 | 36.4 |

## High-pressure angle-dispersive X-ray diffraction

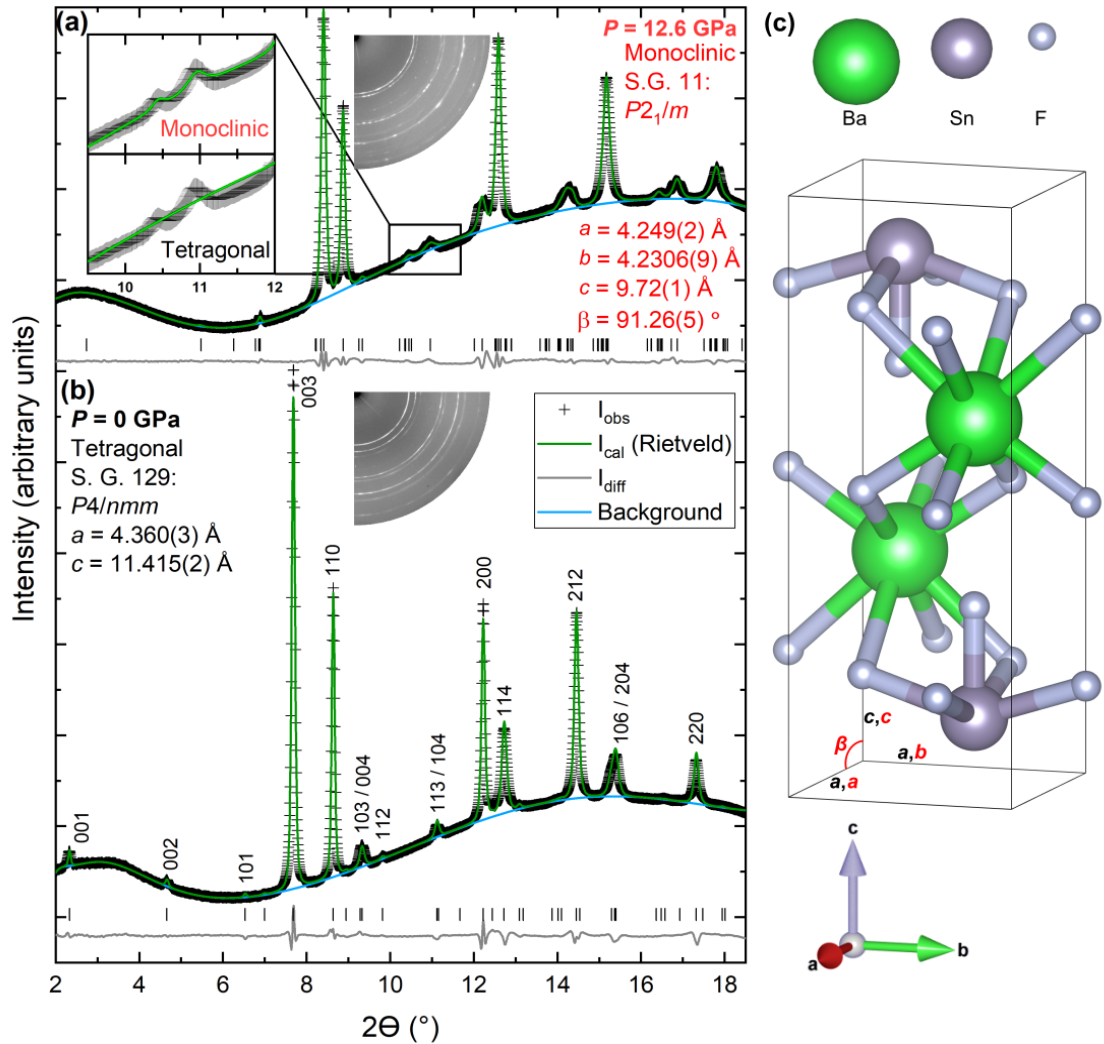

**Figure 4**: Rietveld refinements of X-ray diffraction patterns from BaSnF$_4$ at **(a)** 12.6 GPa and **(b)** 0 GPa. The BaSnF$_4$ crystal structure is shown in **(c)**, where the black (red) labels correspond to the tetragonal (monoclinic) structure. The experimentally observed intensity is shown with black crosses. The calculated intensity of the Rietveld profile is shown in green. The difference between the observed and calculated intensities is shown in grey. The background is shown in blue. The inset shows an enlarged image of the region 9.5 to 12 degrees, where two low intensity reflections are observed which cannot be accounted for by the original tetragonal structure. The two reflections are accounted for by the monoclinic structure as shown in the inset. Vertical tick marks show the positions of the Bragg reflections which are also labelled with their *hkl*s. The insets show the unintegrated 2D diffraction images.

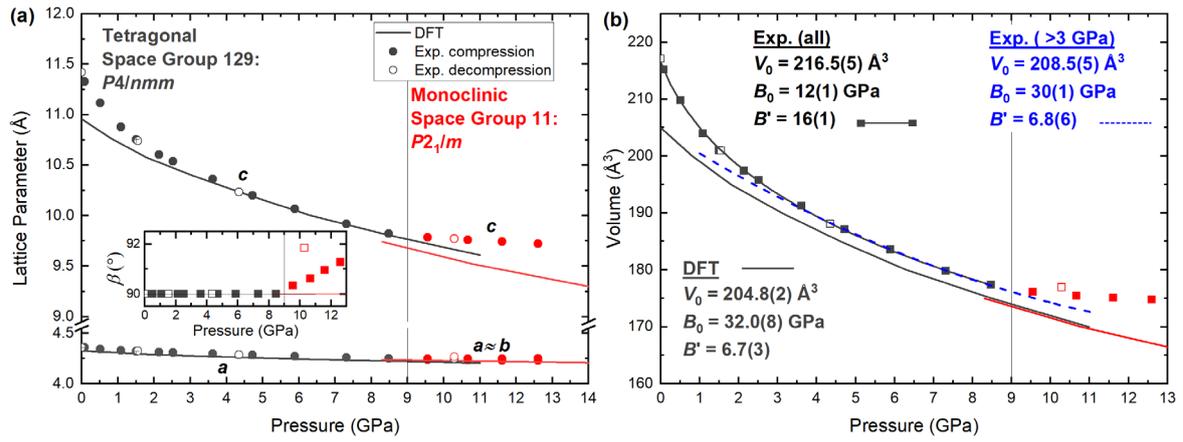

**Figure 5 (a)** Experimental and calculated lattice parameters ($a$, $b$, $c$ and $\beta$) and **(b)** volume for the ambient pressure tetragonal ($P4/nmm$) and high pressure monoclinic ($P2_1/c$) BaSnF$_4$ structures. The tetragonal BaSnF$_4$ data are shown in black. The monoclinic BaSnF$_4$ data are shown in red. Data acquired on (de)compression are shown with (open) solid symbols.

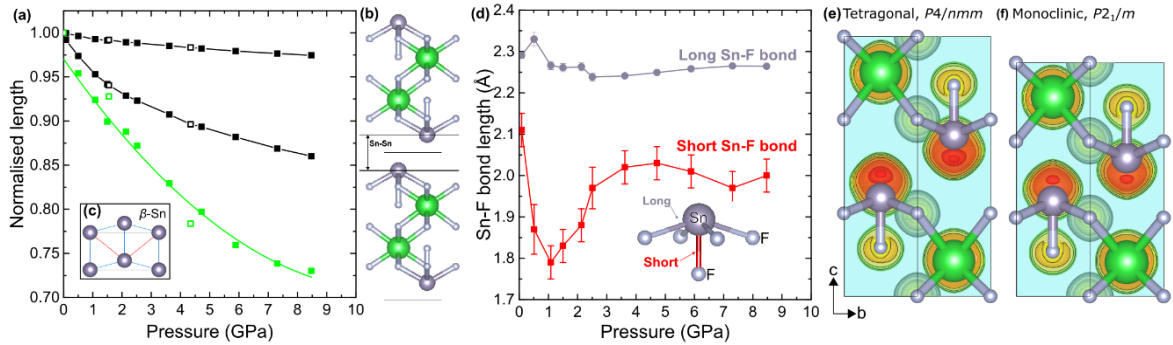

**Figure 6 (a)** Normalised lengths of the unit cell edges ($a$ and $c$) and the Sn-Sn interlayer distance as indicated in **(b)**. **(c)** The $bc$-face of the β-Sn unit cell. Next nearest neighbour distances are shown with red lines. **(d)** The Sn–F bond lengths of the SnF$_5$ pyramid as determined from XRD. **(e-f)** The (1, -1, 0) plane showing the electron localization function (ELF) isosurfaces in the tetragonal and monoclinic-I BaSnF$_4$ phases.

To test the *P*4/*nmm* (ambient pressure) → *P*2$_1$/*m*-I (>10 GPa) hypothesis presented by the DFT calculations, we performed high-pressure angle-dispersive X-ray diffraction measurements on BaSnF$_4$. As starting point, the ambient pressure BaSnF$_4$ XRD pattern was Rietveld refined using the tetragonal *P*4/*nmm* structure of Ref. [26] (see **Figure 4c** for the crystal structure), confirming the published crystal structure and clearly showing a single phase sample – see **Figure 4b**. The difference between the observed and calculated intensities is shown by the grey line. The tetragonal BaSnF$_4$ structure is characterised by BaF$_8$ cubes and SnF$_5$E octahedra, where 'E' represents the Sn lone electron pair (LEP) which points directly along the crystallographic *c*-axis. Notably, the Sn LEP points towards the empty space between two Sn layers.

The *P*4/*nmm* tetragonal structure persisted as the only observable phase in subsequent XRD patterns acquired on compression until *P* = 9.55 GPa, with the patterns exhibiting only shifting of reflections to higher 2Θ values due to compression (i.e. smaller *d*-spacings) within this pressure range. The unit cell parameters determined from the XRD data during the compression phase are shown in **Figure 5a**. The *c*-axis is clearly the most compressible, due to the high compressibility and alignment of the aforementioned Sn LEPs. According to the PASCal principal axis calculator [27], the *c*-axis is the primary compression axis, exhibiting six times greater compressibility than the *a*- or *b*-axes. Over the studied pressure range, the tetragonal *c*-axis contracts by 15.7%, whereas the *a*- and *b*-axes shrink by only 2.8%. The linear compressibility of the tetragonal *a* and *b* axes is 3.16 TPa$^{-1}$, while that of the *c*-axis is 18.05 TPa$^{-1}$. For comparison, the linear compressibility of diamond is approximately 0.75 TPa$^{-1}$.

Using these unit cell parameters, we calculated the pressure-volume equation of state (EOS) for the low-pressure tetragonal phase, shown in **Figure 5b**. Fitting all of the volume data for the tetragonal structure yields a third-order Birch-Murnaghan EOS with a very low bulk modulus, $B_0$ = 12(1) GPa, an exceptionally high bulk modulus pressure derivative, $B'$ = 16(1), and an initial volume of $V_0$ = 216.5(5) Å$^3$.

The very low $B_0$, comparable to that of solidified gases, suggests that this fit is unphysical. This underestimation of $B_0$ arises from the highly nonlinear compressibility of the *c*-axis at pressures below 3 GPa. The high compressibility in this range is linked to the Sn lone electron pair, which points into the Sn-Sn interlayer region. For instance, the Sn-Sn interlayer distance decreases by 15% by a pressure of only *P* = 3 GPa (see **Figures 6a and 6b**). Consequently, the large $B'$ reflects the significant interlayer compressibility due to the rapid reduction in Sn-Sn interlayer distances. Above 3 GPa the Sn-Sn interaction becomes more significant, as indicated by the change in slope of the *c*-axis, reducing the compressibility. At 0 GPa the Sn-Sn interatomic distance is 4.33(2) Å. At 3.6 GPa the Sn-Sn interatomic distance is 3.80(2) Å. This is comparable to the next nearest neighbour Sn-Sn interatomic distance 3.76769(0) Å in β-Sn at ambient pressure [28] shown in **Figure 6c**.

Additionally, the highly nonlinear compressibility of the *c*-axis is underpinned by the pressure-induced evolution of the SnF$_5$ pyramid, as determined via Rietveld refinement of the experimental XRD data. As shown in the inset of **Figure 6d**, the SnF$_5$ unit consists of one short Sn–F bond and four longer, equivalent Sn–F bonds. The short bond is especially significant because it points along the c-axis, (the most compressible direction,) and lies opposite the Sn lone electron pair (LEP). The pressure dependence of the Sn–F bond lengths is illustrated in **Figure 6d**. While the four longer bonds (grey) remain largely unchanged with increasing pressure, the short Sn–F bond (red) exhibits more complex behaviour. It first contracts rapidly from 2.1 Å to 1.8 Å between 0 and 1 GPa, but then, counterintuitively, it lengthens between 1 and 5 GPa. Similar anomalous bond-length increases under pressure have recently been observed in compounds containing $IO_3$ pyramids, attributed to the approach of next-nearest-neighbour atoms [29]. In the case of the $SnF_5$ pyramid, the four fluorine atoms from the adjacent Ba atom (located directly above or below, as shown in **Figures 6e** and **6f**) move closer to the Sn atom under pressure, ultimately reaching a distance of 3.3 Å. This implies that the tetragonal-to-monoclinic phase transition in $BaSnF_4$ is accompanied by a coordination change around Sn, from 5-fold to 5+4-fold.

DFT calculations do not reproduce the highly compressible region below 3 GPa, likely due to the fact that DFT typically underestimates the ambient-pressure volume. In this work, the DFT volume at 0 GPa corresponds to the experimental volume at approximately 1.5 GPa. Shifting the DFT data empirically by this amount further improves the agreement between DFT and experiments.

To address the effect of the high compressibility below 3 GPa on the bulk modulus, a second EOS was calculated, excluding data below 3 GPa. This refinement yields a more reasonable bulk modulus pressure derivative, $B' = 6.8(6)$, with $B_0 = 30(1)$ GPa and $V_0 = 208.5(5)$ Å$^3$. The results from this revised EOS (using data above 3 GPa only) show excellent agreement with the EOS obtained from DFT-derived unit cell parameters: $V_0 = 204.8(2)$ Å$^3$, $B_0 = 32.0(8)$ GPa, and $B' = 6.7(3)$. For comparison, the fluorite $BaF_2$ has a bulk modulus of $B_0 = 57$ GPa, and $B' = 4$ [30]. $BaF_2$ also exhibits a pressure-induced phase transition at the lower pressure of 3 GPa. Therefore, the introduction of Sn not only makes the structure more compressible, due to the empty layers between the Sn atoms, but it also increases the mechanostability.

Above $P = 9.55$ GPa the Bragg reflections between 9.5° and 12° (shown in the inset in **Figure 4**) cannot be accounted for by the original tetragonal $P4/nmm$ symmetry, indicating a phase transition or chemical decomposition. All reflections are accounted for by the monoclinic $P2_1/m$-I symmetry predicted by our DFT calculations as shown in the inset and in the upper panel of **Figure 4**. This tetragonal → monoclinic phase transition constitutes a slight distortion of the original tetragonal structure, wherein the unit cell parameters which were initially $a = b$ in the tetragonal case, become $a \approx b$ in the monoclinic case, with an angle $\beta$ which deviates from 90°. The experimentally determined lattice parameters for the tetragonal $P4/nmm$ and monoclinic $P2_1/m$-I phases, along with the calculated lattice parameters for the tetragonal $P4/nmm$, monoclinic $P2_1/m$-I, and monoclinic $P2_1/m$-II phases, are provided in **Supplementary Tables 1-5**. The corresponding atomic positions are given in **Supplementary Tables 6-10.** While the overall phase evolution and transition pressures are in good agreement between experiment, DFT, Raman, and resistivity measurements, the slightly larger deviation in the experimentally observed β angle compared to DFT may arise from non-hydrostatic stress at high pressures in the experimental sample [31]. Such stress can cause uneven compression along different crystallographic axes, slightly distorting the crystal lattice. This effect is not captured in idealised (hydrostatic) DFT calculations.

The hypothesis of chemical decomposition can be excluded based on two key arguments. Firstly, there is excellent agreement between the calculated and experimental Raman spectra (see later **Figure 7**), whereas decomposition would typically yield new phases, such as $SnF_2$ and $BaF_2$, both known to be Raman-active [32, 33], with distinct vibrational modes. The absence of these characteristic modes in the experimental spectra strongly suggests that $BaSnF_4$ has not decomposed. Secondly, the observed reversibility of the phase transition (as shown by the 'recovered' diffraction pattern in **Supplementary Figure 2**) indicates no permanent chemical change, since pressure-induced decomposition is typically irreversible at ambient temperature [34]. Additionally, the reaction between $SnF_2$ and $BaF_2$ to reform $BaSnF_4$ requires temperatures above 500 °C, while all measurements were conducted at ambient temperature [35].

## High-pressure Raman Spectra

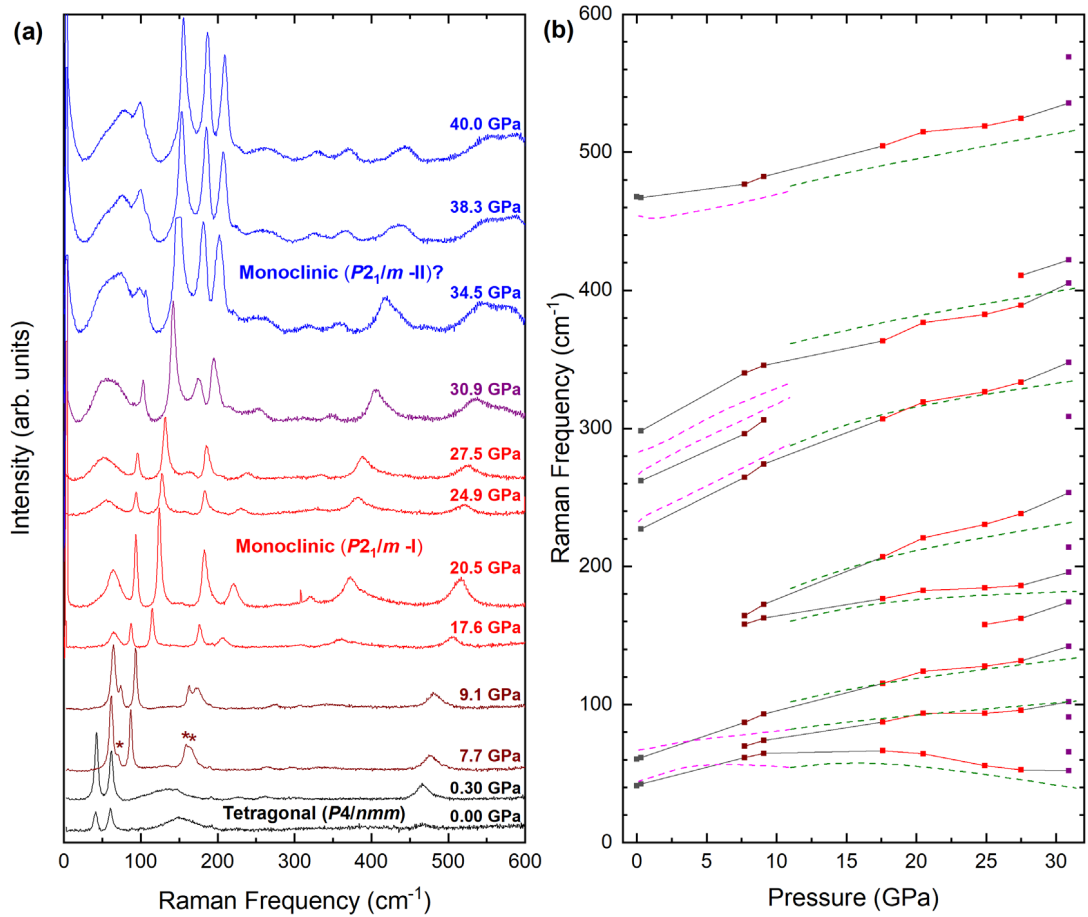

**Figure 7** Experimental and calculated high pressure Raman data for BaSnF$_4$. **(a)** Experimental high pressure Raman spectra. **(b)** Raman frequencies as a function of pressure. The experimental data corresponding to panel (a) are shown with symbols connected by solid lines. Calculated Raman modes are shown with dashed lines, where magenta and green correspond to the tetragonal (*P*4/*nmm*) and monoclinic (*P*2$_1$/*m*-I) phases respectively.

According to group-theoretical analysis, the tetragonal structure of BaSnF$_4$ has twelve Raman-active modes: $4A_{1g} + 2B_{1g} + 6E_g$. In the experimental spectra (see **Figure 7a**), at ambient pressure we detected six Raman modes, and a broad band around 150 cm$^{-1}$ which we do not ascribe to the sample because it was not observable when measuring the sample outside the DAC environment (see **Supplementary Figure 3**). Amongst the observed modes, the strongest are the low frequency modes at 41 cm$^{-1}$ and 60 cm$^{-1}$ and the high frequency mode at 467 cm$^{-1}$. We also detected three week modes at 227, 262, and 298 cm$^{-1}$. According to our DFT calculations the wavenumbers of these modes are 44, 67, 232, 267, 283, and 454 cm$^{-1}$. The agreement is quite good, with wavenumber differences within 10% which is typical for DFT calculations [36]. The agreement is good not only for the value of the wavenumber of the modes at 0 GPa, but also for its pressure dependence as shown in **Figure 7b**. From the DFT calculations, we find that the lowest frequency mode, which corresponds to a mode with Eg symmetry, starts to soften under compression near the transition pressure. This mode involves vibrations of the Sn atoms within the *a-b* plane; *i.e.* within the layers that form the crystal structure. This softening is a precursor to the phase transition and indicates a weakening of the restoring force against the corresponding deformation [37]. This phenomenon is typical of a gradual reduction of symmetry from tetragonal to monoclinic [38] like the one reported here in the present work.

The tetragonal to monoclinic transition is identified in the Raman spectra by the appearance of three Raman modes at a pressure of $P$ = 7.7 GPa. These modes are seen as a shoulder appearing at 70 cm$^{-1}$ and a doublet at 160 cm$^{-1}$, they are all marked with asterisks '*' in **Figure 7a**. According to group-theoretical analysis, the monoclinic structure of $BaSnF_4$ has eighteen Raman-active modes: $12_{Ag} + 6_{Bg}$. In the experiments we detected up to ten Raman modes at 27.5 GPa. The increase in the number of modes is consistent with the phase transition from tetragonal to monoclinic. At 17.6 GPa, there are seven observed modes and they are at 66, 87, 115, 176, 207, 307, 363, and 505 cm$^{-1}$. The calculated frequencies for these modes are 61, 90, 116, 176, 208, 313, 377, and 491 cm-1. As in the case of the low-pressure tetragonal phase, the agreement between calculations and experiments is also good for the high-pressure phase, both for the values of calculated frequencies at 17.6 GPa and for the pressure dependence of the modes observed in experiments (See **Figure 7b**). An interesting observation is that the lowest-frequency mode observed in the HP phase softens under compression. This might be related to the occurrence of a second phase transition as predicted by DFT calculations. The occurrence of such transition is consistent with changes occurring in the Raman spectrum at 30.9 GPa. The predicted HP phase has also eighteen Raman modes, 12Ag + 6Bg. Since the structure of this phase has yet to be confirmed by XRD, we prefer not to compare the pressure dependence of Raman modes of the second HP phase with results from DFT calculations.

**High-pressure Resistivity**

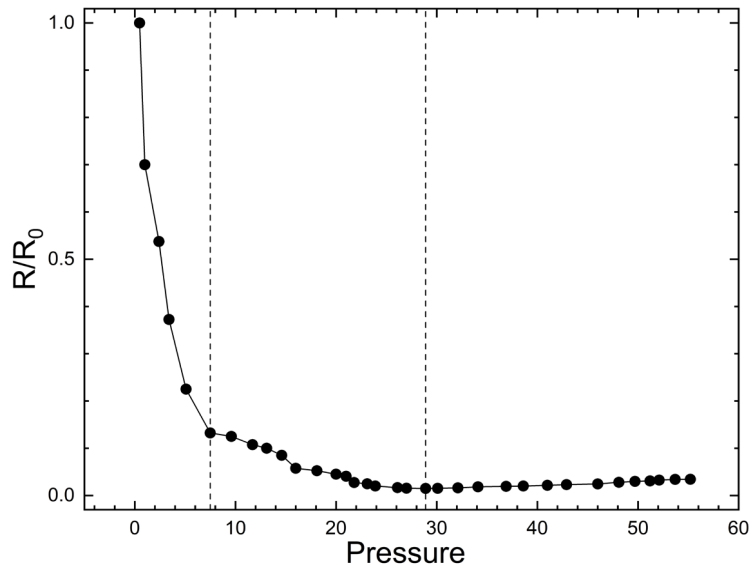

**Figure 8** Results of high-pressure resistivity measurements. Symbols are the results. The line is a guide to the eye. The vertical dashed lines indicate the transition pressures.

**Figure 8** presents the results of high-pressure resistivity measurements. Initially, the resistivity decreases by an order of magnitude from 0 to 7.5 GPa, reaching $R/R_0 = 0.1325$. At 7.5 GPa, there is a change in the slope of the pressure dependence. The potential influence of initial packing effects, such as loosely packed grains or the collapse of voids, on the observed resistivity drop at low pressure has been considered. However, such effects are expected to be negligible above ~1–2 GPa, as any intergranular gaps would be fully closed within this pressure range [39]. Furthermore, the observed linear decrease in resistivity up to 7.5 GPa suggests a systematic, pressure-induced structural modification rather than a packing artefact. This interpretation is supported by complementary evidence from XRD, Raman spectroscopy, and DFT, all of which indicate a phase transition at similar but slightly higher pressures.

Beyond 7.5 GPa, the resistivity continues to decrease slightly with pressure, reaching a minimum of $R/R_0 = 0.015$ at 28.9 GPa, corresponding to a total decrease of two orders of magnitude from 0 GPa. Above 28.9 GPa, the resistivity begins to increase with pressure. The two changes in pressure dependence occur at pressures similar to the observed and predicted transition pressures, providing further evidence of these transitions. The resistivity measurements detect the transitions at slightly lower pressures because the experimental conditions were not hydrostatic.

The decrease in resistivity observed in the low-pressure tetragonal phase cannot be attributed to pressure-induced changes in band-gap energy, as fluorides have a wide band gap [40]. Instead, it is more likely due to a reduction in the activation energy for fluoride migration within the structure, which is 0.3 eV at 0 GPa [41]. A decrease of just 0.05 eV is sufficient to account for the observed change in resistivity. This change is comparable to the effect of increasing the temperature to 500 K [41], which also enhances conductivity in a similar manner to compression. This suggests that pressure creates structural frameworks with connected low-barrier diffusion channels that facilitate fluoride migration.

The slower decrease in resistivity observed in the first high-pressure monoclinic phase (7.5–28.9 GPa) suggests that the ion diffusion mechanism in this pressure range is less sensitive to compression than in the low-pressure phase. While we do not propose a specific diffusion mechanism in this work, similar behaviour has been attributed in previous studies to the preferential alignment of crystallites along the c-axis [41]. Such orientation may impede the reduction in ionic resistivity due to the predominantly two-dimensional nature of fluoride-ion diffusion in $BaSnF_4$. Finally, the increase in resistivity following the second phase transition may be related to partial occupancy of interstitial sites by fluoride ions, which has been associated with a reduction in ionic conductivity [42].

## IV. Conclusions and Discussion

|              | Transition 1 (GPa)<br>$P4/nmm \rightarrow P2_1/m$-I | Transition 2 (GPa)<br>$P2_1/m$-I $\rightarrow P2_1/m$-II |
|--------------|-----------------------------------------------------|-----------------------------------------------------------|
| DFT          | 10.2                                                | 32.4                                                      |
| XRD          | 9.5                                                 | No data                                                   |
| Raman        | 7.7                                                 | 27.5                                                      |
| Resistivity  | 7.5                                                 | 28.9                                                      |

**Table 2** Summary of the phase transition pressures observed in the different diagnostics and calculations in this work.

Density functional theory (DFT) calculations predict two pressure-induced structural phase transitions in the fast-ion conductor $BaSnF_4$. The first, from the ambient tetragonal $P4/nmm$ phase to a monoclinic $P2_1/m$-I structure at ~10 GPa, is confirmed by angle-dispersive X-ray diffraction at room temperature. This transition is also consistent with discontinuities observed in Raman spectra and electrical resistivity. A second transition, from $P2_1/m$-I to a distinct monoclinic phase ($P2_1/m$-II) at ~32 GPa, is supported by pressure-dependent Raman features and a marked change in resistivity behaviour. These findings are summarised in **Table 2**.

The phase transitions identified here indicate that $BaSnF_4$ undergoes meaningful structural reorganisation under pressure with regards to the fluoride-ion environment. Notably, a reduction in resistivity by nearly one order of magnitude is observed in the low-pressure $P2_1/m$-I phase. This suggests enhanced fluoride-ion mobility under compression, consistent with previous observations in single-fluoride materials such as $CaF_2$, where pressure is known to lower the superionic transition temperature and increase ion transport [6].

Compared to other fast-ion fluorides, such as $CaF_2$, $SrF_2$ and $BaF_2$, where pressure-induced structural transitions have been observed [43], $BaSnF_4$ shows a similar trend but in a structurally more complex system due to the layered double-fluoride arrangement. The fact that $BaSnF_4$ maintains a monoclinic symmetry through both high-pressure phases, yet exhibits distinct spectroscopic and transport properties, suggests subtle changes in the local environment of fluoride ions or the mobility pathways. This structural flexibility under compression could make $BaSnF_4$ and related materials promising candidates for pressure-tuned ionic conductors.

These results contribute to a broader understanding of how high pressure affects ion transport in complex fluoride materials. Future work should aim to probe the ionic transport mechanisms in the high-pressure phases directly, possibly using impedance spectroscopy under pressure or molecular dynamics simulations to quantify diffusion pathways. It would also be relevant to explore the $BaSnF_4$ phase diagram at high temperatures.

In summary, this study provides the first experimental investigation of $BaSnF_4$ under pressure, revealing two pressure-induced phase transitions and a pressure-enhanced conductivity response. These findings underscore the potential of pressure as a tool to manipulate structural and transport properties in double-fluoride materials and open up new directions for research in solid-state ionics and high-pressure materials science.

**Supplementary Information**

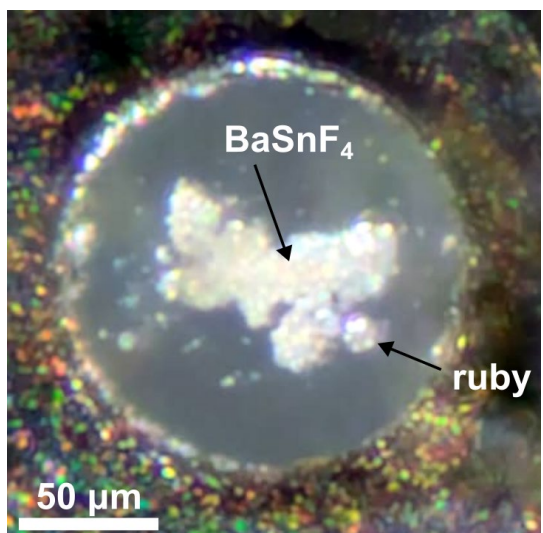

**Supplementary Figure 1** – A typical loading of BaSnF$_4$ in a diamond anvil cell.

**Supplementary Table 1** – Tetragonal *P*4/*nmm* BaSnF$_4$ lattice parameters from Rietveld XRD

| Pressure (GPa) | *a* (Å) | *c* (Å) | Vol (Å$^3$) |
| --- | --- | --- | --- |
| *0.00 outside DAC* | *4.3538(4)* | *11.251(1)* | *213.27(5)* |
| 0.00 | 4.3603(3) | 11.418(1) | 217.08(5) |
| 0.08(5) | 4.3581(3) | 11.327(1) | 215.13(5) |
| 0.51(5) | 4.3441(3) | 11.116(2) | 209.78(9) |
| 1.08(5) | 4.3292(3) | 10.880(1) | 203.90(5) |
| 1.50(5) | 4.3227(3) | 10.755(1) | 200.95(5) |
| 1.56(5) | 4.3248(4) | 10.739(1) | 200.86(5) |
| 2.14(5) | 4.3136(3) | 10.606(1) | 197.34(5) |
| 2.52(5) | 4.3098(3) | 10.539(1) | 195.76(5) |
| 3.62(5) | 4.2956(3) | 10.362(1) | 191.20(5) |
| 4.35(5) | 4.2873(4) | 10.232(1) | 188.08(5) |
| 4.72(5) | 4.2823(3) | 10.202(1) | 187.08(5) |
| 5.89(5) | 4.2702(3) | 10.068(1) | 183.58(5) |
| 7.31(5) | 4.2575(3) | 9.920(1) | 179.81(5) |
| 8.48(5) | 4.2495(3) | 9.821(1) | 177.34(5) |

**Supplementary Table 2** – Tetragonal $P4/nmm$ BaSnF$_4$ lattice parameters from DFT

| Pressure (GPa) | a (Å) | c (Å) | Vol (Å$^3$) |
| --- | --- | --- | --- |
| 0 | 4.325 | 10.958 | 204.9762 |
| 0.8 | 4.311 | 10.764 | 200.0459 |
| 1.8 | 4.294 | 10.576 | 195.0049 |
| 3.1 | 4.276 | 10.393 | 190.0274 |
| 4.6 | 4.258 | 10.204 | 185.0043 |
| 6.3 | 4.241 | 10.007 | 179.9867 |
| 8.5 | 4.224 | 9.807 | 174.9782 |
| 11 | 4.206 | 9.609 | 169.9874 |
| 14.1 | 4.191 | 9.393 | 164.9832 |
| 17.7 | 4.181 | 9.154 | 160.0189 |
| 22.1 | 4.166 | 8.931 | 155.0025 |

**Supplementary Table 3** – Monoclinic $P2_1/m$-I BaSnF$_4$ lattice parameters from XRD

| Pressure (GPa) | a (Å) | b (Å) | c (Å) | β (°) | Vol (Å$^3$) |
| --- | --- | --- | --- | --- | --- |
| 9.56(5) | 4.247(1) | 4.238(1) | 9.783(5) | 90.33(3) | 176.1(2) |
| 10.67(5) | 4.246(1) | 4.233(1) | 9.759(5) | 90.61(3) | 175.4(2) |
| 11.61(5) | 4.247(1) | 4.231(1) | 9.744(7) | 90.95(4) | 175.1(2) |
| 12.60(5) | 4.250(2) | 4.231(1) | 9.72(1) | 91.27(5) | 174.8(3) |

**Supplementary Table 4** – Monoclinic $P2_1/m$-I BaSnF$_4$ lattice parameters from DFT

| Pressure (GPa) | a (Å) | b (Å) | c (Å) | α (°) | β (°) | γ (°) |
|---:|---:|---:|---:|---:|---:|---:|
| 6.2 | 4.252 | 4.252 | 9.955 | 89.972 | 90.002 | 89.992 |
| 8.3 | 4.238 | 4.238 | 9.744 | 89.993 | 89.999 | 89.986 |
| 10.8 | 4.225 | 4.225 | 9.524 | 89.998 | 89.998 | 89.993 |
| 13.9 | 4.211 | 4.211 | 9.306 | 89.992 | 89.998 | 89.996 |
| 17.5 | 4.197 | 4.198 | 9.081 | 89.994 | 89.999 | 89.992 |
| 21.9 | 4.182 | 4.185 | 8.856 | 89.995 | 89.998 | 89.996 |
| 27.2 | 4.166 | 4.196 | 8.581 | 90.002 | 90.000 | 89.911 |
| 32.3 | 4.141 | 4.197 | 8.344 | 90.055 | 90.000 | 89.990 |
| 39.2 | 4.105 | 4.190 | 8.139 | 90.159 | 90.000 | 89.984 |
| 47.6 | 4.064 | 4.167 | 7.972 | 90.291 | 89.999 | 89.992 |
| 57.7 | 4.019 | 4.139 | 7.815 | 90.443 | 90.000 | 89.985 |

**Supplementary Table 5** – Monoclinic $P2_1/m$-II BaSnF$_4$ lattice parameters from DFT

| Pressure (GPa) | a (Å) | b (Å) | c (Å) | α (°) | β (°) | γ (°) |
|---:|---:|---:|---:|---:|---:|---:|
| 22.4 | 5.440 | 4.174 | 6.673 | 90.000 | 98.094 | 90.000 |
| 27.7 | 5.351 | 4.128 | 6.630 | 90.000 | 98.055 | 90.000 |
| 34.3 | 5.258 | 4.085 | 6.585 | 90.000 | 98.149 | 90.000 |
| 42.4 | 5.172 | 4.035 | 6.535 | 90.000 | 98.205 | 90.000 |
| 52.5 | 5.089 | 3.983 | 6.480 | 90.000 | 98.284 | 90.000 |
| 64.9 | 5.008 | 3.929 | 6.421 | 90.000 | 98.377 | 90.000 |
| 80.5 | 4.928 | 3.872 | 6.359 | 90.000 | 98.487 | 90.000 |

**Supplementary Table 6** – Tetragonal *P4/nmm* BaSnF$_4$ atomic positions from XRD at 0.0 GPa.

| Atom | Wyckoff position | Site symmetry | x | y | z | Occupancy |
|---|---|---|---|---|---|---|
| Ba1 | 2c | 4mm | 0.2500 | 0.2500 | 0.8709 | 1.0 |
| Sn2 | 2c | 4mm | 0.2500 | 0.2500 | 0.3670 | 1.0 |
| F1 | 4f | 2mm | 0.7500 | 0.2500 | 0.6948 | 1.0 |
| F2 | 2a | -4m2 | 0.7500 | 0.2500 | 0.0000 | 1.0 |
| F3 | 2c | 4mm | 0.2500 | 0.2500 | 0.1830 | 1.0 |

**Supplementary Table 7** – Tetragonal *P4/nmm* BaSnF$_4$ atomic positions from DFT at 0.0 GPa.

| Atom | Wyckoff position | Site symmetry | x | y | z | Occupancy |
|---|---|---|---|---|---|---|
| Ba1 | 2c | 4mm | 0.2500 | 0.2500 | 0.8640 | 1.0 |
| Sn2 | 2c | 4mm | 0.2500 | 0.2500 | 0.3789 | 1.0 |
| F1 | 4f | 2mm | 0.7500 | 0.2500 | 0.6805 | 1.0 |
| F2 | 2a | -4m2 | 0.7500 | 0.2500 | 0.0000 | 1.0 |
| F3 | 2c | 4mm | 0.2500 | 0.2500 | 0.1875 | 1.0 |

**Supplementary Table 8** – Monoclinic *P2$_1$/m*-I BaSnF$_4$ atomic positions from XRD at 12.6 GPa.

| Atom | Wyckoff position | Site symmetry | x | y | z | Occupancy |
|---|---|---|---|---|---|---|
| Ba1 | 2e | m | 0.7880 | 0.7500 | 0.1330 | 1.0 |
| Sn2 | 2e | m | 0.7760 | 0.7500 | 0.5870 | 1.0 |
| F1 | 2e | m | 0.2444 | 0.7500 | 0.6800 | 1.0 |
| F2 | 2e | m | 0.2496 | 0.7500 | 0.0200 | 1.0 |
| F3 | 2e | m | 0.7452 | 0.7500 | 0.8250 | 1.0 |
| F4 | 2e | m | 0.2536 | 0.7500 | 0.3000 | 1.0 |

**Supplementary Table 9** – Monoclinic *P2$_1$/m*-I BaSnF$_4$ atomic positions from DFT at 11.0 GPa.

| Atom | Wyckoff position | Site symmetry | x | y | z | Occupancy |
|---|---|---|---|---|---|---|
| Ba1 | 2e | m | 0.7512 | 0.7500 | 0.1460 | 1.0 |
| Sn2 | 2e | m | 0.7500 | 0.7500 | 0.5923 | 1.0 |
| F1 | 2e | m | 0.2500 | 0.7500 | 0.6636 | 1.0 |
| F2 | 2e | m | 0.2500 | 0.7491 | 0.9999 | 1.0 |
| F3 | 2e | m | 0.7500 | 0.7521 | 0.8098 | 1.0 |
| F4 | 2e | m | 0.2500 | 0.7484 | 0.3364 | 1.0 |

**Supplementary Table 10** – Monoclinic $P2_1/m$-II BaSnF$_4$ atomic positions from DFT at 34.0 GPa.

| Atom | Wyckoff position | Site symmetry | x | y | z | Occupancy |
|---|---|---|---|---|---|---|
| Ba1 | 2e | m | 0.7367 | 0.2500 | 0.9119 | 1.0 |
| Sn2 | 2e | m | 0.7163 | 0.7500 | 0.4607 | 1.0 |
| F1 | 2e | m | 0.5334 | 0.2500 | 0.2357 | 1.0 |
| F2 | 2e | m | 0.1232 | 0.7500 | 0.4507 | 1.0 |
| F3 | 2e | m | 0.9403 | 0.7500 | 0.7643 | 1.0 |
| F4 | 2e | m | 0.7792 | 0.7500 | 0.1336 | 1.0 |

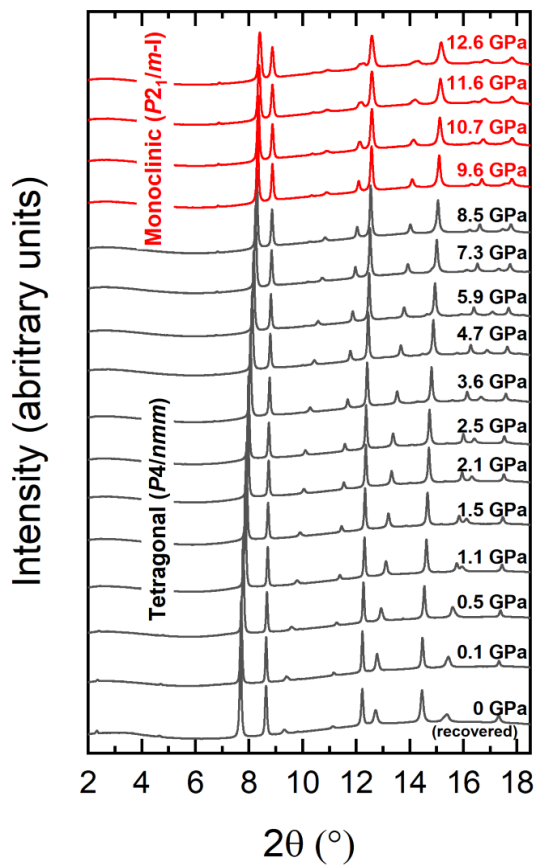

**Supplementary Figure 2** – Waterfall plot of all powder XRD data. Rietveld refinement was performed for each diffraction pattern shown. The corresponding lattice parameters are provided in Supplementary Tables 1 and 3. At 9.6 GPa the tetragonal symmetry undergoes a slight distortion to monoclinic symmetry wherein the β angle deviates from 90°.

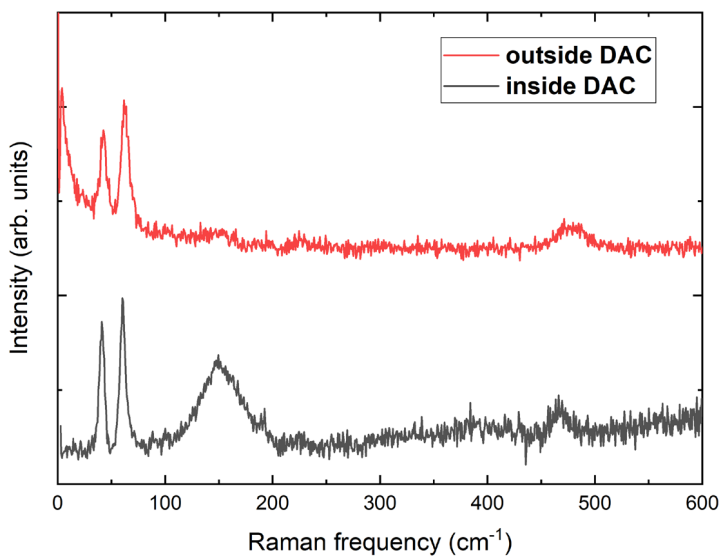

**Supplementary Figure 3** – Raman spectra of $BaSnF_4$ inside (black) and outside (red) of a diamond anvil cell (DAC).

**Acknowledgments**

The authors thank Vitaliy Bilovol and Sergio Ferrari for their contribution to the synthesis of the sample. The authors acknowledge funding from the Spanish Ministerio de Ciencia e Innovación (MICINN) and the Agencia Estatal de Investigación (MCIN/AEI/10.13039/501100011033) under grant No. PID2022-138076NB-C41. This study is part of the Advanced Materials Program supported by MCIU with funding from NextGenerationEU (PRTR-C17.I1) and Generalitat Valenciana. R.T. acknowledges funding from the Generalitat Valenciana for Postdoctoral Fellowship No. CIAPOS/2021/20 and also acknowledges financial support from the Spanish Ministerio de Ciencia e Innovacion through Project No. PID2021-125518NB-I00 financed by MCIN/AEI/10.13039/501100011033. C.C. acknowledges financial support from the Spanish Ministry of Science and Innovation through Grant No. MCIN/AEI/10.13039/501100011033, from the European Union European Regional Development Fund (ERDF) under Project No. PID2023-146623NB-I00, and through the "María de Maeztu" Program for Units of Excellence (CEX2023-001300-M), as well as from the Generalitat de Catalunya (Grant No. 2021SGR-00343). M.P.-A. acknowledges the support of the UKRI Future leaders fellowship MRC-MR/T043733/1. C.P. acknowledges financial support from the Spanish Ministerio de Ciencia e Innovacion through Project No. PID2021-125927NB-C21. L.P. acknowledges financial support from Agencia Nacional de Promoción Científica y Tecnológica under Project No. PICT 201-0364. D. E. acknowledges the support from Generalitat Valenciana under grants CIPROM/2021/075 and MFA/2022/007. The authors thank ALBA for providing beamtime under Experiment No. 2020074389.

**Data Availability**

Data available from the authors upon reasonable request.